\documentclass{article}

\usepackage{style}
\usepackage[utf8]{inputenc} 
\usepackage[T1]{fontenc}    
\usepackage{url}            
\usepackage{booktabs}       
\usepackage{amsfonts}       
\usepackage{amsmath}
\usepackage{nicefrac}       
\usepackage{microtype}      
\usepackage{lipsum}
\usepackage{fancyhdr}       
\usepackage{graphicx}       
\usepackage{glossaries}
\usepackage{siunitx}
\usepackage{multicol}
\usepackage{float}
\usepackage{hyperref}       

\newacronym{ae}{AE}{autoencoder}
\newacronym{gt}{GT}{ground truth}
\newacronym{spc}{SPC}{single-pixel camera}
\newacronym{cmos}{CMOS}{complementary metal-oxide-semiconductor}
\newacronym{mse}{MSE}{mean squared error}
\newacronym{rmse}{RMSE}{root mean squared error}
\newacronym{df}{DF}{datafusion}
\newacronym{dmd}{DMD}{digital micromirror device}
\newacronym{gd}{GD}{gradient descent}
\newacronym{pmt}{PMT}{photo-multiplier tube}
\newacronym{tcspc}{TCSPC}{time-correlated single photon counting}
\newacronym{flim}{FLIM}{fluorescence-lifetime imaging microscopy}
\newacronym{fov}{FOV}{field of view}
\newacronym{dm}{DM}{dichroic mirror}
\newacronym{wf}{WF}{widefield}
\newacronym{slm}{SLM}{structured light modulation}
\newacronym{sam}{SAM}{spectral angle mapper}
\newacronym{roi}{ROI}{region of interest}
\newacronym{cs}{CS}{compressive sensing}
\newacronym{tv}{TV}{total variation}
\newacronym{dof}{DOF}{depth of field}
\newacronym{sh}{$\mathbf{S}$}{Scrambled Hadamard}
\newacronym{psnr}{PSNR}{peak signal-to-noise ratio}
\newacronym{ssim}{SSIM}{structural similarity index measure}
\newacronym{br}{BR}{binary regularization}
\newacronym{led}{LED}{learned encoder-decoder}
\newacronym{ccd}{CCD}{charge-coupled device}
\newacronym{spi}{SPI}{single-pixel imaging}

\newcommand{\s}[1]{\scriptsize{#1}}

\newcommand{\bx}{\mathbf{x}}
\newcommand{\by}{\mathbf{y}}
\newcommand{\bz}{\mathbf{z}}
\newcommand{\beps}{\boldsymbol{\epsilon}}
\newcommand{\bA}{\mathbf{A}}
\newcommand{\bE}{\mathbf{E}}
\newcommand{\bbR}{\mathbb{R}}

\newcommand{\UCL}{Department of Computer Science, University College London, 66-72 Gower St, WC1E6EA London, UK}
\newcommand{\POLIMI}{Dipartimento di Fisica, Politecnico di Milano, Piazza L. da Vinci 32, 20133 Milano, Italy}
\newcommand{\CNR}{Consiglio Nazionale delle Ricerche, Piazza L. da Vinci 32, 20133 Milano, Italy}
\newcommand{\IIT}{Center for Nano Science and Technology, Istituto Italiano di Tecnologia, Via Raffaele Rubattino, 81, 20134 Milano, Italy}
\newcommand{\email}{sc.tudosie.23@ucl.ac.uk}

\title{Learned Single-Pixel Fluorescence Microscopy
}

\author{
  Serban C. Tudosie\ $^{1,*}$ \quad 
  Valerio Gandolfi\ $^{2}$ \quad 
  Shivaprasad Varakkoth\ $^{2}$ \\
  \AND
  Andrea Farina\ $^{3}$ \quad 
  Cosimo D'Andrea\ $^{2,3,4}$ \quad
  Simon Arridge\ $^{1}$
}

\glsdisablehyper

\begin{document}
\maketitle
\NoHyper
\let\thefootnote\relax\footnotetext{$^1$ \UCL}
\let\thefootnote\relax\footnotetext{$^2$ \POLIMI}
\let\thefootnote\relax\footnotetext{$^3$ \CNR}
\let\thefootnote\relax\footnotetext{$^4$ \IIT}
\let\thefootnote\relax\footnotetext{$^*$ \email}
\endNoHyper


\begin{abstract}
Single-pixel imaging has emerged as a key technique in fluorescence microscopy, where fast acquisition and reconstruction are crucial. In this context, images are reconstructed from linearly compressed measurements. In practice, total variation minimisation is still used to reconstruct the image from noisy measurements of the inner product between orthogonal sampling pattern vectors and the original image data.
However, data can be leveraged to learn the measurement vectors and the reconstruction process, thereby enhancing compression, reconstruction quality, and speed. We train an autoencoder through self-supervision to learn an encoder (or measurement matrix) and a decoder. We then test it on physically acquired multispectral and intensity data. During acquisition, the learned encoder becomes part of the physical device.
Our approach can enhance single-pixel imaging in fluorescence microscopy by reducing reconstruction time by two orders of magnitude, achieving superior image quality, and enabling multispectral reconstructions.
Ultimately, learned single-pixel fluorescence microscopy could advance diagnosis and biological research, providing multispectral imaging at a fraction of the cost.
\end{abstract}


\section{Introduction}
\label{sec:introduction}

%
%
Fluorescence microscopy is widely used for cell and tissue imaging and is a fundamental tool in biomedical research~\cite{drummen2012fluorescent}. To image samples in this context, \gls{ccd} and \gls{cmos} cameras are a common choice. Acquiring multispectral information is often desirable but requires costly, slow, and inefficient solutions with tunable spectral filters or point acquisition with a raster scan over the sample~\cite{hiraoka2002multispectral,Zhou1563}. The \gls{spc}~\cite{2008_Duarte_spi} is a new imaging paradigm that can make multispectral acquisitions faster and more efficient, and has been recently applied to multispectral fluorescence microscopy~\cite{2023_Ghezzi_microscope_fusion}. It relies on the principles of \gls{cs}~\cite{2006_Donoho_compressed_sensing}, where compression takes place during observation, thus acquiring a compressed representation of the imaged scene from which the image can be reconstructed.
An \gls{spc} system uses an underdetermined linear measurement matrix and a reconstruction algorithm. The measurement matrix and the reconstruction algorithm can be constructed using the principles of \gls{cs} theory or learned from data~\cite{2018_Higham_DCAN} through self-supervision with an autoencoder. These approaches require learning a physics-constrained encoder (binary, linear, and underdetermined) and a decoder that jointly reconstructs and denoises. For conventional \gls{spc} systems, learned approaches have recently shown advantages in terms of reconstruction quality and speed~\cite{2024_Deng_svd_unet, 2018_Higham_DCAN, 2023_liu_mae_spc}. A crucial aspect of \gls{spc}, in particular, applied to fluorescence microscopy, is reconstructing images as fast as possible to enable real-time imaging for the medical field.

%
%
To our knowledge, there are no studies about the benefit that learned approaches could provide to \gls{spc} systems for fluorescence microscopy.
In addition, it is unclear how essential the training data distribution is for learned \gls{spi}. Thus, it is essential to determine whether a model for single-pixel imaging trained on natural images can be applied effectively in other applications where data availability is limited.

%
%
\begin{figure*}[htbp]
    \centering
    \includegraphics[width=\linewidth]{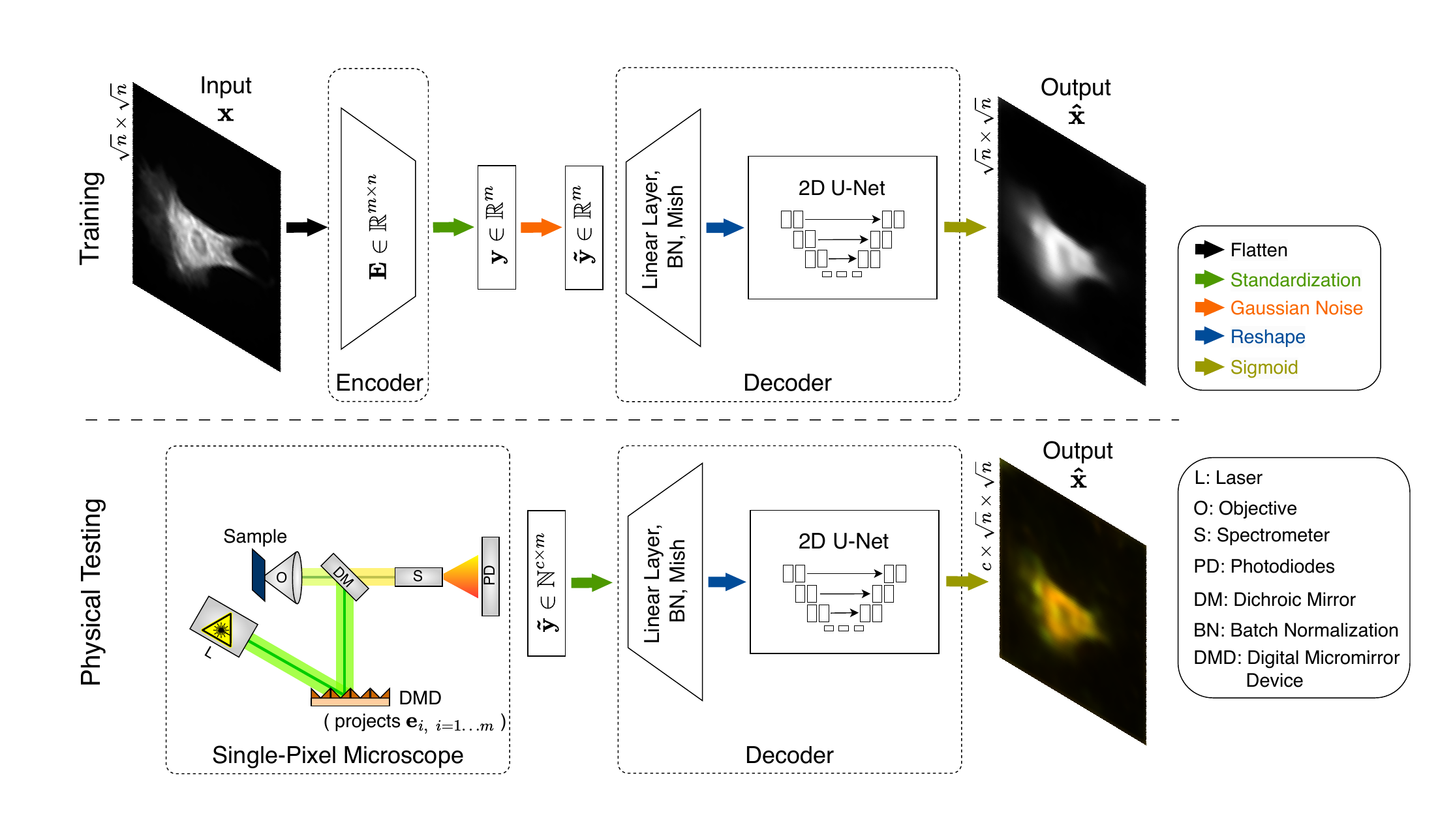}
    \caption{Network architecture. We show the training phase (upper part) and the multispectral physical testing (lower part). At the end of the training phase, $\bE$ becomes binary. At test time, the decoder can be used over multiple spectral channels, using the batch dimension to create an output image with $c$ channels; or, the spectral channels can be integrated so that the decoder can be tested on single-channel measurements. The microscope box reports the setup of the single-pixel multispectral fluorescence microscope.}
    \label{fig:schema}
\end{figure*}
These questions motivate the current study, which aims to develop an autoencoder that is easily trainable for \gls{spc} microscopy setups, works in the multispectral setting, and can achieve real-time reconstructions. We showcase how it behaves at different noise levels, compression percentages, and training datasets. We then compare it with standard \gls{cs} approaches. We additionally use the measurement matrix learned by the model's encoder for experimental measurements with an \gls{spc}-based multispectral fluorescence microscope setup and then reconstruct with its learned decoder, comparing the results to standard \gls{cs} approaches. 

\section{Methods}
\label{sec:method}
\subsection{Single-Pixel Camera}
The \gls{spc}~\cite{2008_Duarte_spi} is an imaging technique that exploits structured light illumination (or detection) and a single-pixel photodetector~\cite{2020_Gibson_12yearsreview}. The \gls{spc} fluorescence microscope used in our physical measurements generates structured light illumination through a \gls{dmd}. A laser illuminates the \gls{dmd}, producing a series of spatial light patterns by turning individual mirrors on or off. Each pattern creates a single measurement with the photodetector – a dot product between the scene and the pattern. The multispectral acquisition is obtained by coupling the \gls{spc} with a spectrometer and an array of single-pixel photodetectors, such that each photodetector captures a different spectral band. In the multispectral setting, the same pattern is used for each band.

The forward procedure is linear and can be expressed as $\by = \bA \bx$, where $\by \in \mathbb{R}^{m \times 1}$ is the measurement vector, $\bx \in \bbR^{n \times 1}$ is the image, $\bA \in \{0,1\}^{m \times n}$ is the measurement matrix (we will refer to this matrix as "patterns" or "encoder" as they perform the same function); $m, n \in \mathbb{N}$. Practically, each row in $\bA$ gets reshaped into a $\sqrt{n} \times \sqrt{n}$ pattern, which the \gls{dmd} can expose. $\bA$ must be binary, i.e. $\in \{0, 1\}^{m \times n}$ for physical requirements of \glspl{dmd}. The goal is to retrieve $\bx$. In a noiseless setting, if $\bA$ is invertible, retrieving $\bx$ is trivial, $\bx = \bA^{-1}\by$. However, in physical \gls{spc} systems, noise is always present due to the discrete nature of photons and the fluctuations in circuits. This results in a Poisson-Gaussian noise model 
\begin{equation}
    \mathbf{\tilde{y}} = \gamma \bz + \beps,
    \text{where}\ 
    \bz \sim \mathcal{P}(\frac{\bA\bx}{\gamma})\ 
    \text{and}\  
    \beps \sim \mathcal{N}(0, \sigma^2 \mathbf{I}).
    \label{eq:noise_model}
\end{equation}
Additionally, most applications aim to have a fast acquisition time, i.e. reducing $m$ as much as possible, which makes the problem underdetermined. 

\Gls{cs} plays a key role in the \gls{spc} setting. \gls{cs} states that retrieving a signal by randomly sampling it, even at a sub-Nyquist rate, is possible if the signal is sparse in some domain, e.g. Fourier. Thus, we can retrieve $\bx$ at $m \ll n$ measurements, but $\bA$ must be incoherent to the basis in which the signal is sparse and must respect the restricted isometry property~\cite{2007_Baraniuk_cs_requirements}. \gls{sh} patterns~\cite{2016_Huynh_sh} are a common choice for $\bA$ since they are orthogonal, pseudo-random, and \gls{sh} $\in \{-1, 1\}^{m \times n}$, i.e. more adequate for \gls{cs}. To directly adapt them to the optical setup, we will consider \gls{sh} patterns where $-1$ entries are mapped to $0$, and we will refer to these patterns again as \gls{sh}. This choice comes at the loss of orthogonality. Thus, in practice, in a noisy, compressed setting ($m<n$), a common procedure is to approximately retrieve $\bx$ by \gls{tv} regularisation, solving an ROF~\cite{1992_ROF} model with \gls{sh} measurement matrix:  
\begin{equation}
\displaystyle \min_{\bx}\ \frac{\mu}{2}||\text{\gls{sh}}\bx - \by||^2_2 + \text{TV}(\bx).
\end{equation}
TVAL3~\cite{2013_Li_tval3} is an efficient algorithm to approximate $\bx$ for \gls{spc}, often used in fluorescence microscopy~\cite{2023_Ghezzi_microscope_fusion}. Alternatively, it is possible to learn from data an underdetermined $\bA$ together with a decoding function $D_{\theta}$ parametrised by $\theta$, through a self-supervised autoencoder~\cite{2018_Higham_DCAN}. Then, physically measure with a learned $\bA$ and reconstruct with $D_{\theta}$ at test time. We will refer to the learned $\bA$ matrices as $\bE$.
\begin{figure*}[tp]
    \centering
    \includegraphics[width=0.99\linewidth]{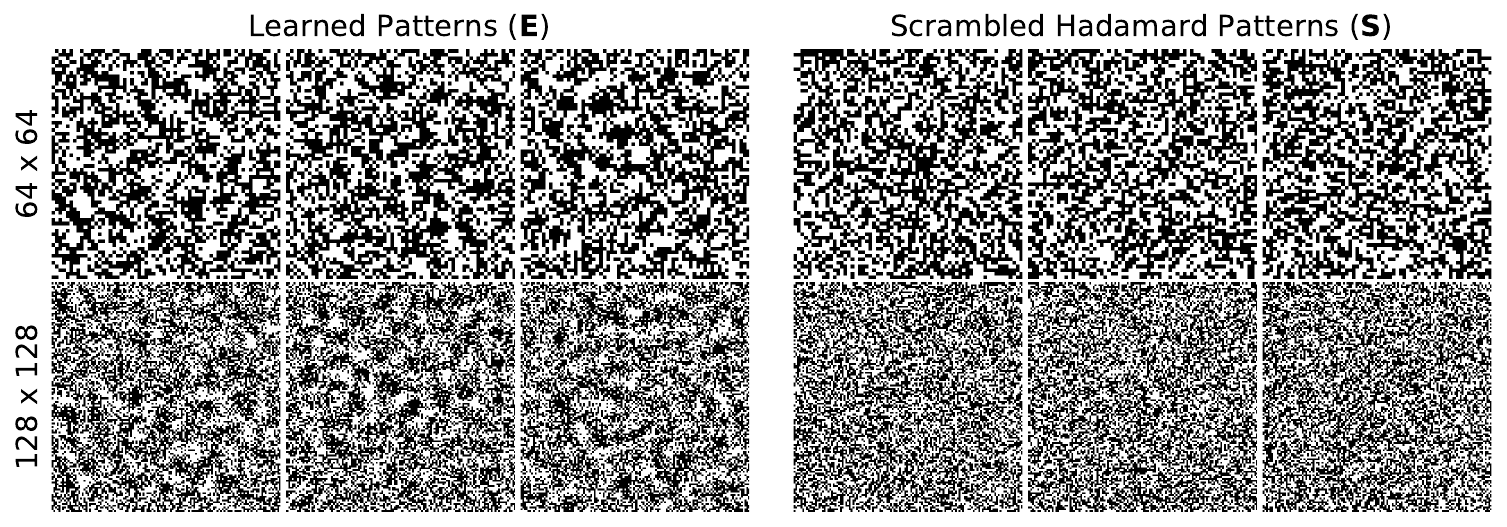}
    \caption{Patterns from the learned encoder (left) and \gls{sh} patterns (right) for $64 \times 64$ (top) and $128 \times 128$ (bottom). We show three random rows from the measurement matrices reshaped in their 2D representation. These 2D patterns correspond to the structured light emitted on the sample.}
    \label{fig:patterns}
\end{figure*}

\subsection{Our Approach}
We use an autoencoder to jointly learn a binary linear underdetermined $\bE$ and a reconstruction network $D_{\theta}$ (Fig.~\ref{fig:schema}). We take inspiration from~\cite{2018_Higham_DCAN} and~\cite{2024_Deng_svd_unet}, where an $\bE \in \{-1,1\}^{m \times n}$ is learned from data, while the decoder learns to denoise and reconstruct the image from a noisy latent space (simulated measurements). 
However, in contrast to previous works, we modify the losses and the reconstruction network. In more detail, we change the binarisation penalty term ($\mathcal{L_B}$) so that $\bE\in\{0,1\}^{m \times n}$ at the end of the training, instead of $\bE\in\{-1,1\}^{m \times n}$. The latter would require $2 \times m$ measurements (first the positive patterns and then negative ones) since an optical setup with one DMD can encode only 0 (dark pixel) and 1 (bright pixel).  
We thus use the following non-convex binarisation penalty on $\bE$: 
\begin{equation}
    \mathcal{L}_B(\bE) = \frac{\lambda}{mn}\sum_{i=1}^{m}\sum_{j=1}^{n}(e_{ij} - 1)^2 e_{ij}^2.
\end{equation}
The parameter $\lambda$ is dynamic. At each network update, starting from $0$, we increase $\lambda$ by $\frac{1}{TS}$, where $TS$ is the total number of training steps. When $\mathcal{L}_B(\bE)$ stops improving, we stop updating $\bE$ and continue training to further adapt $D_\theta$ to the completely binarised $\bE$. We initialise $\bE$ from a beta distribution $e_{ij} \sim \mathcal{B}(\alpha, \beta)$ with $\alpha = \beta = 0.7$, which empirically provides faster training. Then, we use a combination of L1 loss ($\mathcal{L}_1(\bx,\hat{\bx}) := || \bx - \hat{\bx}||_1$) and SSIM loss ($\mathcal{L}_{\text{SSIM}}(\bx,\hat{\bx}) := 1 - \text{SSIM}(\bx,\hat{\bx})$) for our data fidelity term since their combination has been proven ideal for image restoration tasks~\cite{2017_Zhao_nvidia_imgrestore}. In summary, we solve the following problem: 
\begin{equation}
    \min_{\bE,\theta} 
    \frac{1}{|\mathcal{D}|}\sum_{i=1}^{|\mathcal{D}|}
    \Bigl[
    w_1 \overbrace{\mathcal{L}_1(\bx^{(i)},\hat{\bx}^{(i)})}^{\text{\scriptsize Intensity}}\ +\ 
    w_2 \overbrace{\mathcal{L}_{\text{SSIM}}(\bx^{(i)},\hat{\bx}^{(i)})}^{\text{\scriptsize Structure}}
    \Bigr]\ +\ 
    w_3 \overbrace{\mathcal{L}_B(\bE)}^{\text{\scriptsize Binarisation}}
    ,
\end{equation}
where $\hat{\bx}$ is the reconstruction $D_{\theta}(\bE\bx + \beps)$, $w_1, w_2, w_3 \in \mathbb{R}$ are weights that we set to $0.2$, $0.8$, and $16$, respectively, and $|\mathcal{D}|$ is the cardinality of the dataset. We only focus on additive Gaussian noise for simplicity.
We standardise the measurements $\by = \bE\bx$ (instance normalization~\cite{2016_Ulyanov_instancenorm} without learnable parameters). Standardisation is crucial for adapting the decoder to real measurements, as different acquisitions might have different means that depend on the power of the laser, exposure time, and the sample itself. To the standardised measurements, we add Gaussian noise to learn a decoder that also performs denoising. 

For $D_\theta$, we employ a linear layer followed by a 2D U-Net~\cite{2015_unet}. We use Mish~\cite{2019_Misra_mish} to introduce non-linearities and Adam~\cite{2014_Kingma_adam} to optimise. Note that the network has non-linearities only in $D_\theta$ as the forward process is purely linear to match the physics and design of the single-pixel camera microscope. Fig.~\ref{fig:schema} shows the network diagram at training and physical testing times.
In this paper, we refer to our approach as \gls{led}.

\subsection{Datasets}
We employ a reduced version of CytoImageNet~\cite{2021_cytoimagenet}, a single-channel microscopy dataset with cellular images. In particular, we are interested in image sizes of $64 \times 64$ and $128 \times 128$. In structured illumination microscopy, larger image sizes are unfeasible, since higher pattern resolutions reach the diffraction limit for projected patterns. For training, validation, and testing, we randomly select, respectively, $ 100,000$, $ 10,000$, and $ 1,000$ images, each at a size of $128 \times 128$. We call this dataset Cyto128. Then, we downscale this dataset to $64 \times 64$ to create Cyto64. We use a small test set since \gls{tv}-based approaches require orders of magnitude more time to retrieve the image. 

To demonstrate the generalisation capability of the network, we also use $100,000$ images from the STL10 dataset~\cite{2011_stl} to train a \gls{led} model (STL-LED). This dataset contains natural images, e.g. horses and cars. We then test it on cellular data.

To show the applicability of \gls{led} on real-world measurements for fluorescence microscopy, we experimentally acquire a test set on a biological sample. We use bovine pulmonary artery endothelial cells (FluoCells™ Prepared Slide \#1, Invitrogen™). From this slide, we collect $3$ \glspl{fov} of \qty{150}{\micro\metre} $\times$ \qty{150}{\micro\metre}. The cells are stained with multiple fluorescent dyes and will show different emission spectra in separate parts of the cell under multispectral imaging. Particularly, cell membranes should emit at \qty{512}{\nano\meter}, while mitochondria at \qty{599}{\nano\meter}. Additionally, we couple the microscope with an extra output to a \gls{cmos} camera to provide a ground truth intensity image. Thus, each entry in the test set is made of a \gls{cmos} image (spatial ground truth), compressed multispectral single-pixel measurements with learned $\bE$ patterns, compressed multispectral single-pixel measurements with \gls{sh} patterns, and uncompressed multispectral single-pixel measurements with \gls{sh} patterns (to reconstruct the spectral ground truth image).

\section{Experiments}
\label{sec:experiments}
\begin{figure}[htbp]
    \centering
    \includegraphics[width=0.55\linewidth]{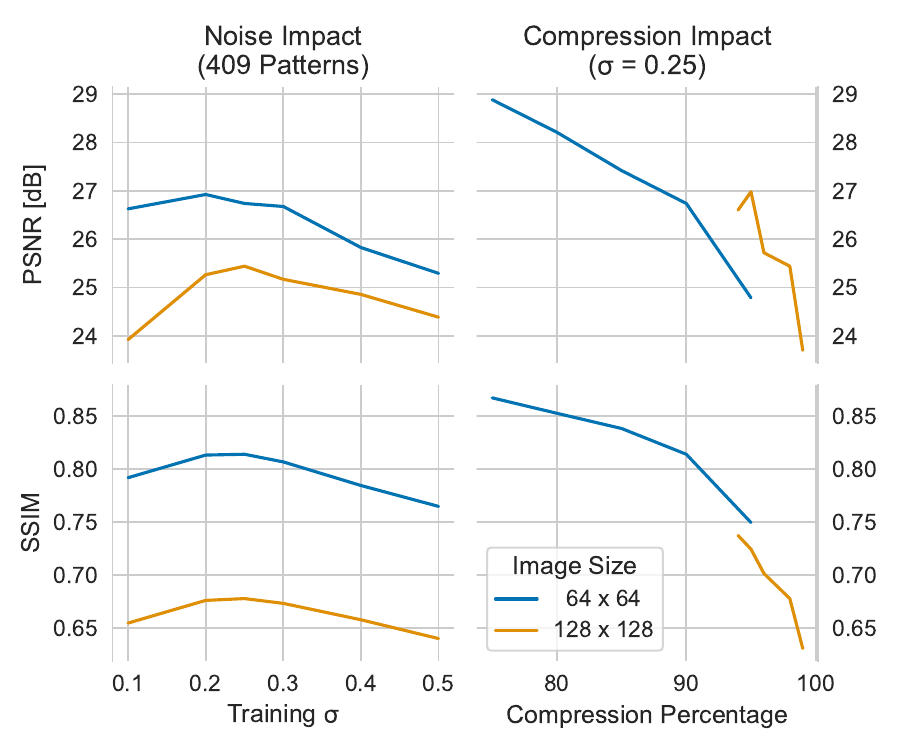}
    \caption{Noise (left) and compression (right) impact on \gls{led}'s performance on Cyto64 and Cyto128 test sets. We show PSNR and SSIM to quantify performance. To evaluate the noise impact on the model, we fix the number of patterns at $409$; similarly, to evaluate the compression, we fix the noise at a standard deviation $\sigma = 0.25$.}
    \label{fig:compression_noise}
\end{figure}

\subsection{Noise-Compression Model Behaviour}
\begin{figure*}[htbp]
    \centering
    \includegraphics[width=\linewidth]{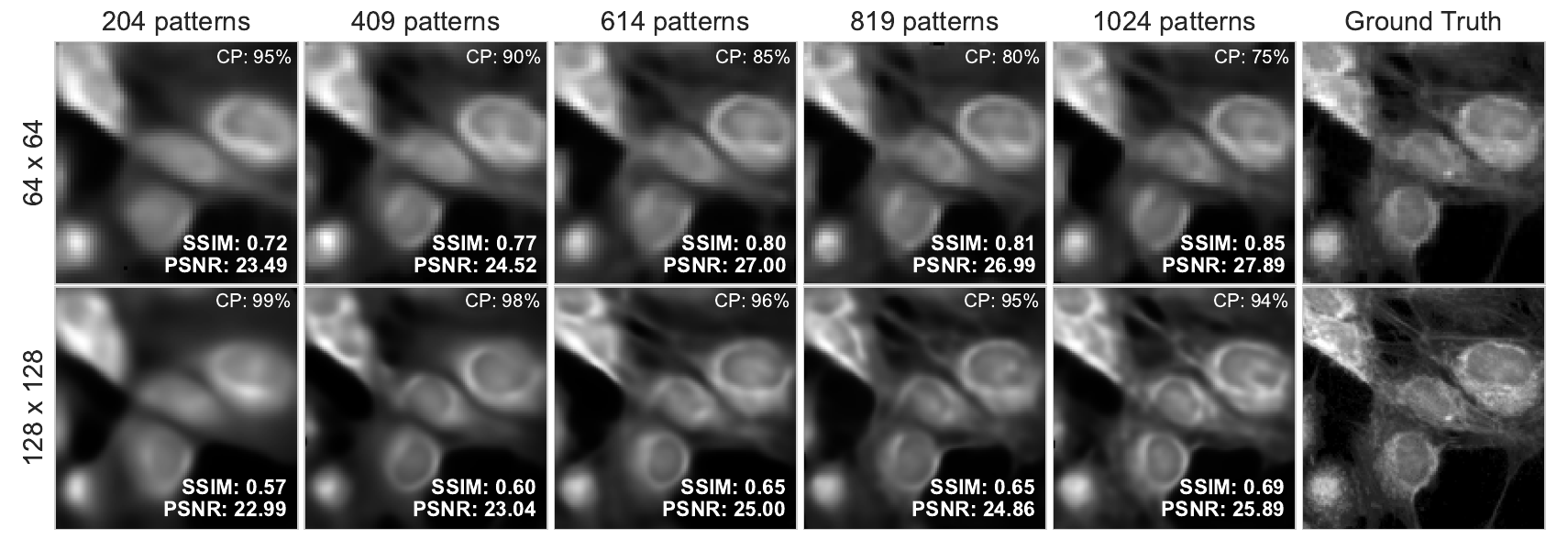}
    \caption{Reconstructions on an image from the test sets Cyto64 (first row) and Cyto128 (second row) with \gls{led} models trained with different compression levels. For each model, we report \gls{ssim}, \gls{psnr}, and the compression percentage CP. The last column shows the corresponding ground truth.}
    \label{fig:compression_reconstruction}
\end{figure*}

We train \gls{led} models on Cyto64 and Cyto128 training sets at $6$ different training noise levels (standard deviation $\sigma$ of additive Gaussian noise), and then, at test time, we add noise with $\sigma=0.25$ (Fig.~\ref{fig:compression_noise} left). We also train at $5$ different compression percentages $CP := 100\ (1 - m / n)$ and a fixed noise level (Fig.~\ref{fig:compression_noise} right). We test the trained models on Cyto64 and Cyto128 test sets. We observe that the models depend on the training noise level in a non-monotonic fashion. In other words, a model trained at $\sigma=0.5$ and tested at $\sigma=0.25$ will perform worse than a model trained at $\sigma=0.25$ and tested at $\sigma=0.25$. This suggests that for excessively large training $\sigma$, the models stop benefiting from the addition of noise, and it may be beneficial to train with different noise $\sigma$s to increase robustness. Compression plays an important role on test \gls{ssim} and \gls{psnr}. More compression results in faster acquisitions and reduced photobleaching, albeit at the expense of lower-quality reconstructions. Between acquiring at $CP=95$ to $CP=75$ the $64 \times 64$ models have a $0.1$ increase in \gls{ssim}. We show a reconstruction with \gls{led} models from the Cyto test sets in Fig.~\ref{fig:compression_reconstruction}. \gls{led} models maintain lower frequencies when constrained to high compression (Fig.~\ref{fig:compression_reconstruction} 204 patterns) and gradually include higher frequencies for lower compression constraints (Fig.~\ref{fig:compression_reconstruction} 1024 patterns); similarly to how an inverse Fourier transform from the $k$-highest coefficients would act at different thresholds $k$.

\subsection{Performance on Simulated Measurements}

\textbf{Considered Methods.} 
\begin{table*}[bp]
\centering
\begin{tabular}{c|ccccccc}
\toprule
Image Size & Metric & SH-TVAL3 & LE-TVAL3 & SH-LD & DCAN & STL-LED & LED \\
\midrule
                  & SSIM      & 0.39 \s{± 0.10} & 0.51 \s{± 0.10} & 0.60 \s{± 0.13} & 0.71 \s{± 0.10} & 0.74 \s{± 0.13} & \textbf{0.82 \s{± 0.09}} \\
$64 \times 64$    & PSNR [dB] & 15.59 \s{± 3.87} & 18.3 \s{± 3.21} & 21.38 \s{± 3.20} & 23.37 \s{± 2.89} & 22.19 \s{± 3.70} & \textbf{26.61 \s{± 3.12}} \\
                  & Time [ms] & 418 \s{± 310} & 662 \s{± 467} & 11 \s{± 8 }& \textbf{4 \s{± 3}} & 11 \s{± 5} & 14 \s{± 7 }\\
\midrule
                   & SSIM      & 0.23 \s{± 0.08} & 0.44 \s{± 0.12} & 0.50 \s{± 0.14} & 0.62 \s{± 0.13} & 0.63 \s{± 0.15} & \textbf{0.71 \s{± 0.13}} \\
$128 \times 128$   & PSNR [dB] & 12.53 \s{± 2.87} & 17.13 \s{± 3.59} & 20.22 \s{± 3.16} & 23.6 \s{± 3.30} & 21.46 \s{± 3.37} & \textbf{25.67 \s{± 3.74}} \\
                   & Time [ms] & 469 \s{± 524} & 636 \s{± 1095} & 23 \s{± 13} & \textbf{9 \s{± 4}}& 22 \s{± 9 }& 26 \s{± 14} \\
\bottomrule
\end{tabular}
\caption{
Comparison of methods on the Cyto64 and Cyto128 test sets, reporting mean ± standard deviation of \gls{ssim}, \gls{psnr}, and reconstruction time. In this case, we use $409$ patterns for the  $64 \times 64$ comparisons and $614$ for the $128 \times 128$ ones with a fixed $\sigma = 0.25$.
}
\vspace{-0.05in}
\label{tab:comparison}
\end{table*}

We now compare different methods for acquiring the compressed measurements and reconstructing the image. We experiment with simulated measurements from images from Cyto64 and Cyto128 test sets. In particular, we consider the following methods: SH-TVAL3, LE-TVAL3, SH-LD, DCAN, STL-LED, and \gls{led}.
SH-TVAL3 consists of measuring with \gls{sh} patterns and reconstructing with TVAL3. LE-TVAL3, instead, first learns the patterns with a \gls{led} model, then measures with the learned $\bE$ but reconstructs with TVAL3. SH-LD learns the decoder by training a \gls{led} model with a frozen measurement matrix \gls{sh} patterns. DCAN is the model from \cite{2018_Higham_DCAN} trained with our loss function, weight initialisation, and latent space standardisation. STL-LED is a \gls{led} model trained on natural images from the STL10 dataset instead of microscopy images; all the other \gls{led}-based methods and DCAN are trained on Cyto64 and Cyto128 datasets.
We add noise with $\sigma=0.25$ for all methods and fix the compression to $409$ patterns ($CP = 90$) for $64 \times 64$ images and $614$ patterns ($CP = 96$) for $128 \times 128$ images.
TVAL3-based reconstruction requires setting $2$ parameters of the algorithm, $\mu$ and $\beta$; which we set in the following way: $\mu = 2^9, \beta = 2^4$ for SH-TVAL3 and $\mu = 2^7, \beta = 2^4$ for LE-TVAL3.

\textbf{Methods Comparison.} 
Table~\ref{tab:comparison} shows a comparison between the considered methods in terms of \gls{ssim}, \gls{psnr}, and time required to reconstruct. We can observe that learning either the measurement matrix or the decoder benefits the reconstruction quality. Additionally, using a learned decoder instead of solving the inverse problem for each image takes significantly less reconstruction time. The smallest decoder (DCAN) requires, on average, \qty{4}{\ms} to reconstruct $64 \times 64$ images, while the TVAL3 reconstructions can take $100$ times longer. \gls{led} improves in both \gls{ssim} and \gls{psnr} with respect to the DCAN approach. Lastly, we show that \gls{led} can generalise by training it on the STL10 dataset (STL-LED) and testing it on Cyto64 and Cyto128 (Table~\ref{tab:comparison}). TVAL3 reconstruction quality depends extensively on the choice of the internal parameters $\mu$ and $\beta$, making it difficult to select the optimal parameters over a large and diverse set of images. In contrast, with learned approaches, the hyperparameters can be chosen using the validation set, making the reconstruction quality at test time more consistent and parameter-free. 
We compute the reconstructions for the comparisons on the test sets on an Apple Silicon M3 CPU.

\noindent\textbf{Patterns Comparison.} 
In Fig.~\ref{fig:patterns} we show the learned patterns and the \gls{sh} patterns. Learned patterns exhibit more structure and lower frequencies. During training, we leave the patterns unconstrained (besides the binarisation term in the loss). However, after training, the learned patterns reach a fill factor: $\frac{1}{mn}\sum_{i=1}^{m}\sum_{j=1}^{n}e_{ij}$, of $0.5$ which corresponds to the fill factor of \gls{sh} patterns. In addition, the learned ones exhibit quasi-orthogonal properties, similarly to \gls{sh} patterns. Thus, most of the singular values of $\bE$ are constant (making each pattern contribution to the measurements maximally informative).

\section{Physical Testing}
\label{sec:results}
\begin{figure*}[htbp]
    \centering
    \includegraphics[width=\linewidth]{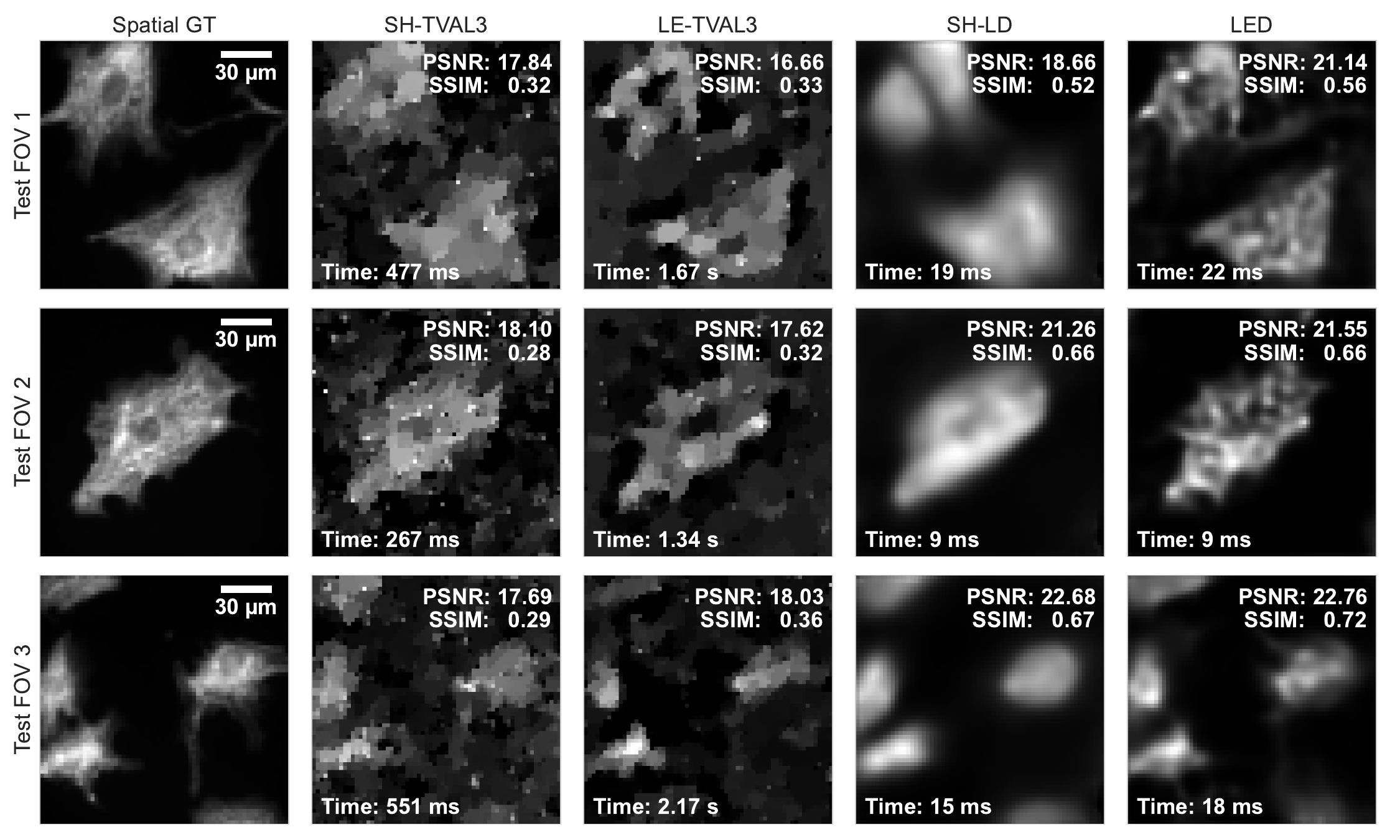}
    \caption{Comparison between single-pixel intensity imaging approaches on a physical microscope at $90\%$ compression. Each row shows a different test sample. The first column represents the spatial intensity \gls{gt}. The following columns show the different approaches. For each reconstruction, we report \gls{psnr}, \gls{ssim} (top right), and reconstruction time (bottom left).}
    \label{fig:results_intensity}
\end{figure*}

\subsection{Intensity Reconstructions}
First, we work with fluorescence intensity measurements (single channel measurements without spectral information), as in the model. We obtain the measurements by summing the data from the multispectral \gls{spc} microscope along the axis corresponding to the spectrum.
We find that \gls{led} models can significantly improve the reconstruction quality and speed on a physical single-pixel microscope (Fig.~\ref{fig:results_intensity}). We compare four methods: SH-TVAL3, LE-TVAL3, SH-LD, and \gls{led} for three \glspl{fov}.  We demonstrate that prior learning of the patterns can aid the TVAL3 reconstruction, improving cellular structure (Fig.~\ref{fig:results_intensity} LE-TVAL3), e.g. in FOV 1 and FOV 2, the cell membranes maintain a closer structure to the \gls{gt} image. Similarly, learning the decoder with \gls{sh}  encoding (SH-LD) reduces background noise, improving all metrics. Finally, \gls{led} benefits from both effects. Time is significantly reduced when using the learned decoder because instead of solving a minimisation problem for each acquisition, it just applies the learned mapping $D_\theta$. We remark that reducing reconstruction time is crucial in translating current single-pixel microscopy into real-time imaging.

\subsection{Multispectral Reconstructions}
\begin{figure*}[htbp]
    \centering
    \includegraphics[width=\linewidth]{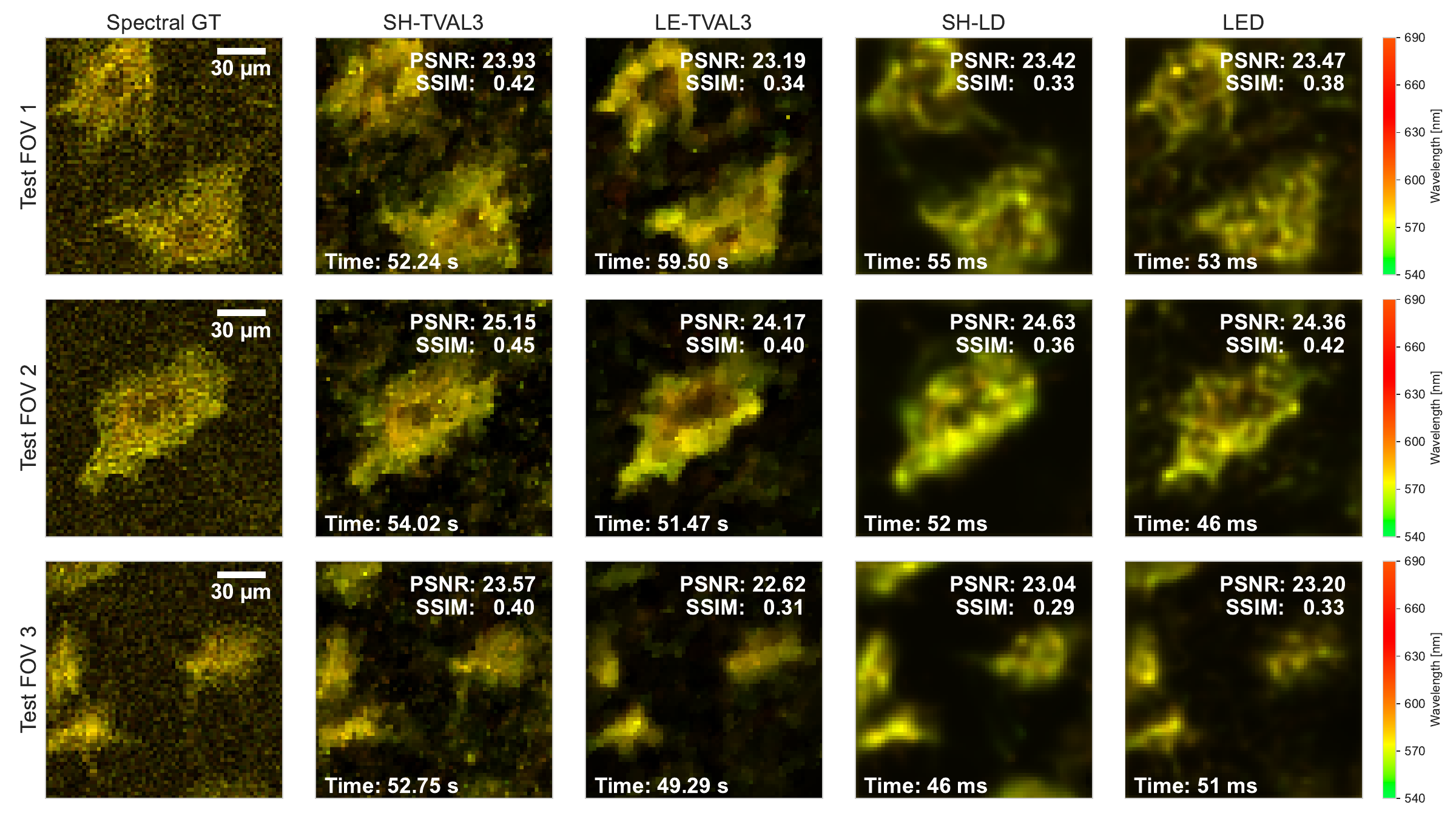}
    \caption{Comparison between single-pixel multispectral imaging approaches on the physical microscope. Each spectral channel undergoes $90\%$ compression. Each row shows a different test sample. The first column represents the \gls{gt}. The following columns show the different approaches. The colour bar shows the detector wavelength range. For convenience of plotting, the $16$-channel spectral images are transformed to sRGB using the CIE standard colour-matching functions \cite{1932_smith_cie}. For each reconstruction, we report the \gls{psnr} over the entire cube, the mean \gls{ssim} of the $16$ channels (top right), and reconstruction time (bottom left).}
    \label{fig:results_spectrum}
\end{figure*}
We additionally find that \gls{led} models can be applied to multispectral imaging (Fig.~\ref{fig:results_spectrum}). In this context, the measurements are multispectral, i.e. each pattern produces $c$ measurements as shown in Fig.~\ref{fig:schema}, at $c$ different wavelengths. We thus reconstruct multispectral images by applying the four previous methods to each channel separately (Fig.~\ref{fig:results_spectrum}). However, models that reconstruct with the learned decoder standardise the latent space. To account for different channel intensities (i.e. the spectral shape), we save the normalised channel means of the measurements before applying standardisation and then, after the reconstruction, we multiply each of the channels by the saved mean. Since the images from the \gls{cmos} camera are not multispectral, we acquire the entire basis of \gls{sh} patterns and use the inverse \gls{sh} matrix to retrieve an image for each channel of the single-pixel measurements. We consider this inversion the spectral \gls{gt}. Without further denoising, this \gls{gt} will look corrupted by Poisson-Gaussian noise. However, in this case, we focus on showing the capability of reconstructing multispectral images at a reduced computational time. Fig.~\ref{fig:results_spectrum} shows the enormous reduction in reconstruction time for $64 \times 64$ images with $16$ spectral channels, while maintaining similar image quality. The cells are correctly reproduced, with the yellow-green colour highlighting the membrane and the orange colour in the inner part highlighting the mitochondria.

\section{Conclusion}

%
%
A crucial aspect of single-pixel camera fluorescence microscopy systems is to make image reconstruction fast and reliable. We demonstrate that it is possible to apply a learning paradigm to such systems, thereby improving upon classical \gls{cs} approaches in terms of reconstruction speed and image quality. 
We showcase the comparison on a physical setup for intensity and multispectral reconstructions, thus demonstrating a real-world application of our model. 
Additionally, we show that the model exhibits particular generalisation capabilities, as by training on natural images, we can still reconstruct cellular images with a similar performance to the model trained on cellular images. This opens new avenues in learned \gls{spi} applied to scenarios where data is scarce, e.g. endoscopy.

%
%
The current limitation of learned \gls{spi} is the unmatched noise modelling with respect to the real-world case. This can be seen directly from the performance of the models on the Cyto64 test set ($0.82$ \gls{ssim}) compared to the reconstructions from the real-world measurements ($0.65$ \gls{ssim}). Here, we have only considered the additive Gaussian noise case; however, a more realistic case in \glspl{spc} is Poisson-Gaussian noise~\ref{eq:noise_model}, which should be considered in future works. 
Additionally, \gls{spc} fluorescence microscopes can be coupled with an additional device (a time-correlated single photon counter \cite{2007_becker_multispectral}) to produce an extra temporal dimension in the measurement for each pattern – sampling fluorescence decay over time. The lifetime of the decay can help investigate metabolic processes \cite{2018_Sanchez_metabolic, 2023_Karrobi_metabolic}. We currently limit the application of learned \gls{spi} to multispectral and intensity measurements.
Lastly, an apparent limitation could appear to be the reduced image size. However, \gls{spc} microscopy systems similar to the one shown in Fig.~\ref{fig:schema} are usually coupled with a high-resolution \gls{cmos} camera. The intensity \gls{cmos} image and the multispectral single-pixel reconstruction can be combined through data fusion~\cite{2021_Soldevila_df, 2023_Ghezzi_microscope_fusion}, providing a high-resolution multispectral image, without the need for a multispectral or hyperspectral \gls{cmos} sensor, and the potential to work in the infrared range and beyond.

%
%
Future work could enhance the robustness of learned \gls{spi} in microscopy by attempting to denoise measurements affected by Poisson-Gaussian noise during training. The validity and application of learned \gls{spi} approaches in the lifetime imaging setting are yet to be confirmed. 
The reconstruction quality and potentially the training speed may benefit from different binarisation strategies, such as the straight-through estimator~\cite{2013_bengio_ste}, but more work and comparisons are needed. Similarly, different overall training strategies and different decoding networks might be beneficial, such as VAEs~\cite{Kingma_Welling_2013_VAEs} and unrolled architectures~\cite{Wu_2018_CS_unrolling}.
Additionally, the standardisation of the latent space could be surpassed by making the decoder scale-invariant as explained in~\cite{2020_drunet, 2019_scale_inv_bias_free}, which could improve multispectral reconstructions.

%
%
Our findings pave the way for real-time multispectral fluorescence microscopy. Indeed, the short time ($50$ ms) needed for reconstruction is compatible with acquisitions at a frame rate greater than 10 fps. This enables fluid visualisation that could find applications in the intraoperative field. 
%
%
In conclusion, we have demonstrated that learned approaches are necessary for more efficient compression, faster reconstruction times, and improved reconstruction quality, thereby extending the applicability of single-pixel fluorescence microscopes to real-time scenarios.

\section*{Funding}
The authors acknowledge financial support by the UK government’s Horizon Europe funding guarantee (grant number EP/X030733/1) and the European Union (GA 101072354).
Views and opinions expressed are, however, those of the author(s) only and do not necessarily reflect those of the European Union, the UK government, or the European Research Executive Agency. Neither the European Union nor the granting authority can be held responsible for them.

\bibliographystyle{unsrt}  
\bibliography{references}

\end{document}